\documentclass[twocolumn,showpacs,preprintnumbers,amsmath,amssymb]{revtex4}
\usepackage{graphicx}
\usepackage{epsfig}
\usepackage{dcolumn}
\usepackage{color}
\usepackage{bm}
\usepackage{natbib}

\definecolor{grey}{rgb}{.65,.65,.65}

\begin{document}
\title{Far-infrared electrodynamics of thin superconducting NbN film in magnetic field}

\author{M. \v{S}indler$^{1}$,
 R. Tesa\v{r}$^{1,2}$,
 J. Kol\'a\v{c}ek$^1$,
 P. Szab\'o$^3$,
 P.~Samuely$^3$,
 V.~Ha\v{s}kov\'{a}$^3$,
 C.~Kadlec$^1$,
 F.~Kadlec$^1$,
 and  P.~Ku\v{z}el$^1$
}

\affiliation{
$^1$Institute of Physics, Academy of Sciences, Cukrovarnick\'{a} 10/112,
16200 Prague 6, Czech Republic}

\affiliation{
$^2$ Faculty of Mathematics and Physics, Charles University,
Ke Karlovu 3, 12116 Prague 2, Czech Republic}

\affiliation{ $^3$ Centre of Low Temperature Physics, Institute of Experimental Physics,
Slovak Academy of Sciences and Faculty of Science, P.J. \v{S}af\'{a}rik University,
Ko\v{s}ice, Slovakia }

\begin{abstract}
We studied a thin superconducting NbN film in magnetic fields up to 8 T above the zero-temperature limit by means of
time-domain terahertz and scanning tunneling spectroscopies in order to understand the 
vortex response. Scanning tunneling spectroscopy was used to determine the optical gap and the upper critical field of the sample.  The obtained values were subsequently used to fit the terahertz complex conductivity spectra in the magnetic field
in the Faraday geometry above the zero temperature limit. These spectra are best described in terms of the
Coffey-Clem self-consistent solution of a modified London equation in the flux creep
regime.
\end{abstract}


\maketitle
\section {Introduction}
Terahertz-range electrodynamics of superconductors is governed by the response of Cooper pairs and
thermally activated quasiparticles~\cite{Tinkham1956a,Tinkham1956b,Tinkham1958,Uwe}. The key parameter is the optical gap
$2\Delta$ which represents the energy necessary for breaking Cooper
pairs~\cite{BCS_original_paper}. For classical BCS-like superconductors the value of
the optical gap ranges from tenths to units of meV which corresponds to the high-frequency
part of the microwave range and to the terahertz (THz) region.

In a magnetic field $B$ lying between the lower and upper critical fields
($B_{c1}<B<B_{c2}$), the type-II superconductors, such as NbN, enter in the so-called
Abrikosov or mixed state. In this state, the magnetic field penetrates the superconductor through
 cylindrical regions carrying  a quantized magnetic flux  $\Phi_0$, so-called vortices:
the normal state is locally restored in vortex cores.
The vortex lattice has a large impact on the high-frequency electrodynamic response of superconductors
 not only because the material becomes inhomogeneous, but also because a vortex oscillatory
 motion inevitably contributes to both the inductive and the dissipative 
 parts of the electrodynamic response \cite{Coffey-Clem_model_1991,Brandt1991,Dulcic1993}.
While the vortex state has been studied quite thoroughly in the microwave range
 \cite {PompeoSilva2008,Rosenblum1964,Janjusevic2006,GRmodel}, there are only 
 few studies in the far-infrared range lying in the vicinity of the optical gap \cite{Ikebe2009,Xi2013}.
 The conclusions of these studies, however, do not provide a simple unambiguous 
 picture:  Ikebe et al.~\cite{Ikebe2009} explained 
their experiments on classical NbN superconductor  by the Coffey-Clem model
 within the dissipative flux-flow regime, while Xi et al.~\cite{Xi2013} 
 argued that the vortices are strongly pinned 
and the contribution from the vortex motion can be neglected.

The aim of this paper is to clarify the issue by studying a superconducting NbN  film in the Faraday geometry
in a broad temperature range between $T=0\,K$ and a temperature slightly above the critical temperature $T_c$.
A number of important parameters of the studied NbN film were determined by supplemental scanning tunneling spectroscopy (STS) experiments,
 which enabled us to test various theoretical models with a minimum of free parameters.

\section{Experiment}
NbN was chosen as a typical representative of classical BCS type-II superconductors. Our
sample was a 11.5~nm thick NbN film deposited on a highly resistive Si substrate. The
nominal critical temperature determined by dc conductivity measurements was 
$T_c=11.5\,$K.

A custom-made spectrometer was used for time-domain terahertz spectroscopy (TDTS) measurements
in the transmission geometry. Broadband THz pulses were generated using a Ti:sapphire
femtosecond laser and a commercial large-area semiconductor interdigitated emitter
(TeraSED, GigaOptics). The sample was placed in an Oxford Instruments Spectromag He-bath
cryostat with mylar windows and a superconducting coil, enabling measurements down to
$T=2\,\mbox{K}$. The distance of the sample from the built-in temperature sensor was of
several cm; therefore, at low temperatures, the real value of the temperature at the
sample was several K higher than the readout on the temperature controller. The Faraday
configuration was used with an external static magnetic field $B\le 7\,\mbox{T}$ parallel
with the wave vector of the THz radiation. The transmitted pulses were detected by
electro-optic sampling \cite{Nahata99} in a 1\,mm thick $\langle110\rangle$ ZnTe crystal.

The experiments consisted of two consecutive measurements: one of a signal wave form
$E_s(t)$ using a sample consisting of a NbN film on substrate, and one of a reference wave
form $E_r(t)$ with a bare Si substrate. These wave forms were time-windowed such that they
contained only the direct pass of the THz beam through the thick substrate
\cite{Duvillaret96}. The ratio of the Fourier transforms of the time-domain wave forms
provided the complex THz transmittance spectrum of the film: $t(\omega) =
E_s(\omega)/E_r(\omega)$. The complex refractive index $\tilde{n}=n+i\kappa$ of the NbN film was
then determined by numerically inverting the expression~\cite{conventionnote,kadlecch}:
\begin{eqnarray}
&&\!\!\!t(\omega)=  \\
&&\!\!\!\frac{2 \tilde{n} (n_s+1)\exp[ik_0(\tilde{n}-1)d]\exp[ik_0(n_s-1)(d_s-d_r)]}{(1+\tilde{n})(\tilde{n}+n_s)+(1-\tilde{n})(\tilde{n}-n_s)\exp[2ik_0 \tilde{n}d]},
\nonumber
\end{eqnarray}
where $n_s$ is the refractive index of the substrate and $k_0=\omega/c$ with $c$ denoting
the speed of light in vacuum. The film thickness is denoted by $d$; $d_s$ and $d_r$
are the thicknesses of the substrate under the film and of the reference (bare)
substrate, respectively. The complex conductivity spectra $\tilde{\sigma}(\omega)$ of the
thin film were directly evaluated from $\tilde{n}(\omega)$ as $\tilde{\sigma}=-i\omega\varepsilon_0\tilde{n}^2$. 
Inverting Eq. (1) may offer multiple solutions in thick samples; however,  in the thin film regime where $|k_0\tilde{n}d|\ll 1$, the determination of the refractive index of the film is
 unambiguous. Moreover, in the thin film approximation the transmission function depends directly on the product
$\tilde{\sigma}d$, which means that a possible error in the film thickness determination
only scales the amplitude of the measured conductivity spectra and does 
not change their shape.

STS measurements were performed in a homebuilt low-temperature scanning tunneling microscope (STM) 
inserted in a commercial Janis SSV cryomagnetic system with $^3$He refrigerator and controlled by Nanotec’s Dulcinea SPM electronics. 
The sample and the tip were cooled down
to 0.7~K and magnetic fields up to 8~T was applied perpendicularly to the NbN film. 
Sharp gold tip was prepared in-situ by cold welding of the Au wire to the gold spot and subsequent retracting the wire \cite{tsamuely}. 
The tip was then scanned over the NbN sample. 
Bias voltage was applied to the tip, while the sample was grounded. 
Au tip enabled a formation of the
N-I-S tunnel junction, where N stayed for the gold tip (normal state), I is the insulating barrier of vacuum, resp. surface oxides and S stands for the superconducting NbN sample. 

The tunneling spectrum, i.e.  the differential conductance versus the voltage 
dependence obtained by STS, represents a convolution of the local 
density of states of both electrodes comprising the tunneling junction. 
Since the normal tip features a constant density of states near the Fermi energy, the 
differential conductance  $dI/dV$ versus voltage  of a N-I-S junction 
reflects the local energy-dependent superconducting density of states $N_S(E)$ smeared 
by $\approx \pm 2k_BT$ in energy at temperature $T$, where $k_B$ is the Boltzmann constant:
\begin{eqnarray}
&&\!\!\!\frac{dI}{dV}(V) \propto \int\limits_{-\infty}^{\infty}N_S(E)\frac{\partial       f}{\partial E}(E+eV)dE,
\end{eqnarray}
where $f$ is the Fermi-Dirac distribution. The tunneling probability which also 
enters into the expression of the tunneling current is taken as a constant at low voltages where the superconducting energy gap is scanned. 
For the BCS superconductor $N_S(E) = E/\sqrt{E^2-{\Delta}^2}$, where $\Delta$ is the superconducting energy gap.
Consequently, in the low temperature limit ($k_BT << \Delta$), the differential conductance measures directly the superconducting density of states $N_S$  \cite{tinkham}.  

\section{Results}
First, our experimental results in zero magnetic field are evaluated, then 
our efforts are extended to include magnetic field measurements in 
the Faraday geometry. 

\subsection{Zero magnetic field}

STM measurements have revealed that an oxidized layer covers the whole surface of our NbN
film, thus preventing us from a detailed topographic investigation of the sample surface.
Semenov et al.~\cite{Semenov2009} estimated the thickness of oxidized layers on the surface of
NbN films exposed to the air and found that layers approximately 1~nm thick are built. In order to perform STS measurements on such samples
a force was applied to the Au tip allowing it to penetrate through the oxide layer. Such 
measurements were performed at several  places of the sample surface and
 revealed a slight variation  of the superconducting gap.

\begin{figure}[!h]
 \center
\includegraphics [width=0.47\textwidth]{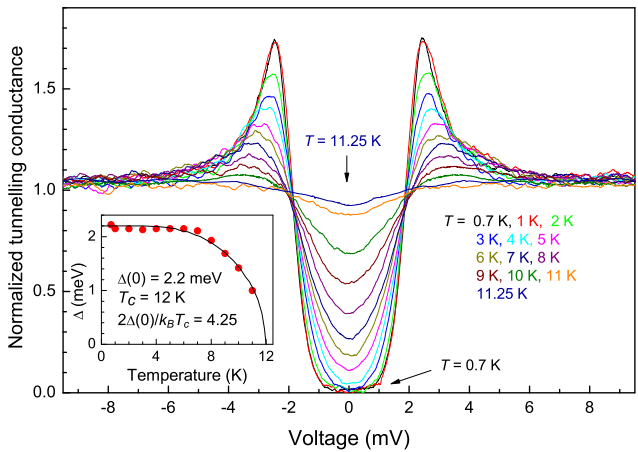}
\caption{Differential tunneling conductance of the NbN-Au tunnel junction measured at
several temperatures between 0.7 and 12.5 K. The inset shows the temperature dependence
of the energy gap $\Delta(T)$ in a zero magnetic field together with the prediction of
the BCS model (full line).} \label{STMspectra}
\end{figure}

Typical STM spectra taken at indicated temperatures above 0.7 K are displayed in Fig. 1. 
The dependences of differential tunneling conductance on the bias voltage were normalized to the spectrum taken in the normal state at $T=12.5$ K. The value of the
superconducting gap $2\Delta(0)$ corresponds roughly to the distance of the peaks in the
tunneling conductance taken at the lowest temperature. The temperature dependence of the gap was determined by fitting
the spectra by Eq.~2 taking into account the thermal broadening. Moreover, in the BCS density of states an extra smearing parameter $\Gamma$ has been incorporated 
by posting the complex energy $E'=E+i\Gamma$ in the density of states expression~\cite{dynes}. This enables to account phenomenologically for a broadening of the spectra of unknown origin, here probably caused by the nanoscale inhomogeneity of the sample. From fit at 0.7 K we obtained a relatively small value of $\Gamma = 0.12 \Delta$ which was then temperature independent. 
The obtained temperature dependence of the superconducting gap is shown in the inset of
Fig.~\ref{STMspectra} together with the prediction of the BCS theory~\cite{tinkham} adjusted for $2\Delta(0)/k_B T_c=4.25$ and a  reasonable agreement is found.

The superconducting energy gap $\Delta(0)$ and the local transition temperature $T_c$ have been checked at several locations on the sample surface. The local values of the gap $\Delta(0)$ estimated from the fits of the low temperature spectra to Eq.~2 varied between 2.1 and 2.3 meV. The local $T_c$ has been independently determined from the temperature dependence of the energy gap $\Delta(T)$ and from its extrapolation to zero. The resulting $T_c$ varied between 11.5 and 12 K. These results points to a moderate inhomogeneity of the NbN film. Since the thickness of our sample corresponds only to a few coherence lengths, we conclude that this inhomogeneity is rather in plane than along the depths of the film. Indeed, the  coherence length $\xi(0)=\sqrt{\Phi_0/2\pi B_{c2}(0)}$ estimated from
the upper critical field
$B_{c2}(0)\approx 26\,\mathrm{T}$ (see below) amounts to 3.5\,nm.
Thus, our STS measurements determine a ratio of $2\Delta(0)/k_B T_c=4.25-4.5$ in our NbN films suggesting a strong coupling superconductivity. 

For steady-state TDTS experiments
a bare 404\,$\mu\mathrm{m}$ thick highly resistive Si substrate was used as a reference. First, the normal-state properties of the NbN film were
determined just above the superconducting phase transition. Using the Drude model
($\tilde{\sigma}_n=\sigma_n(0)/(1-i\omega \tau)$) the value of DC conductivity $\sigma_n(0)=0.45\,\mu\Omega^{-1} \mathrm{m}^{-1}$ was obtained while
 the relaxation time $\tau$ could not be reliably determined. We estimated $\tau < 15$~fs, the best fit yielded $\tau=8$~fs ($1/(2\pi\tau)=20$~THz).

\begin{figure}
 \center
\includegraphics [width=0.47\textwidth]{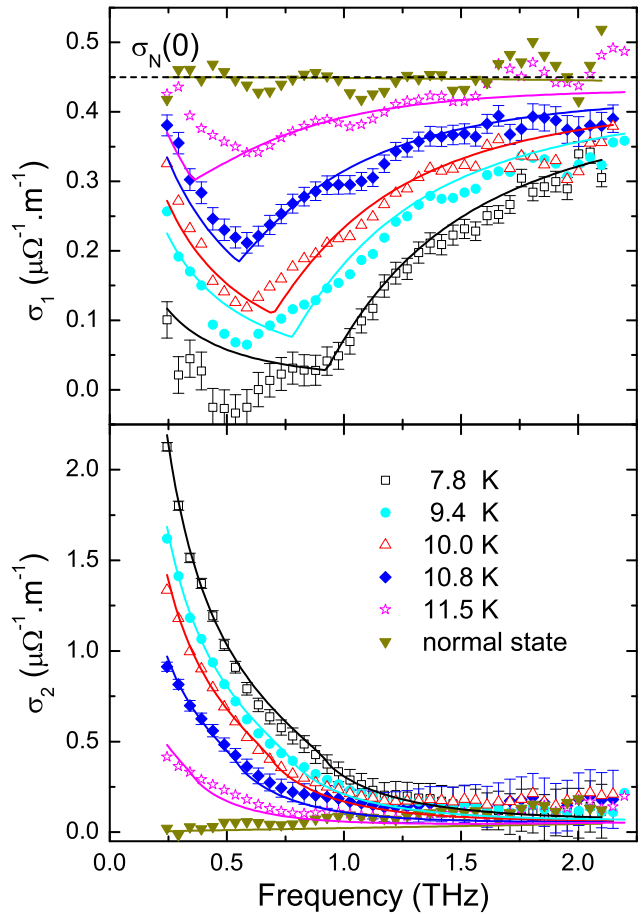}
\caption{Real ($\sigma_1$) and imaginary ($\sigma_2$) parts of the conductivity of NbN
thin film in zero magnetic field at several temperatures. The following values were used
in fits using the Zimmermann model \cite{Zimmmermann} (solid lines): $\sigma_n(0)=0.45\, \mu\Omega^{-1}$.m$^{-1}$, $\tau=8$ fs, $\Delta(0)=2.2$~meV and $T_c=12$~K.}
\label{Zimmerman}
\end{figure}

As pointed out above, the temperature sensor in the cryostat was not in 
a direct
contact with the sample and some temperature gradient appeared. This did not 
influence the
measurements in the normal state since the normal-state properties are only weakly
temperature dependent. Below $T_c$ we fitted the measured complex conductivity at each
temperature by the Zimmerman model \cite{Zimmmermann} with the temperature as the only free
parameter (see Fig.~\ref{Zimmerman}). We observe in Fig.~\ref{Zimmerman} that $\sigma_2$
exhibits a typical $1/\omega$ dependence and that it decreases with increasing
temperature since the density of condensate decreases. The sharp-edge feature in
$\sigma_1$ is a characteristic fingerprint of the superconducting gap; it becomes less
pronounced at higher temperatures due to an increasing number of quasiparticles.
Close to $T_c$,  $\sigma_1$ continuously transforms into $\sigma_{1n}$. Note that a careful
evaluation of the experimental errors (error bars in Fig.~\ref{Zimmerman}) shows that
$\sigma_2$ is determined more precisely than $\sigma_1$ and that especially below the
superconducting gap (where $\sigma_1$ acquires quite small values) the real part of the
conductivity suffers from large relative uncertainties. This is an important statement
which will be used for fitting the experimental spectra obtained in the magnetic field.

\subsection{Non-zero magnetic field}

\begin{figure}
 \center
\includegraphics [width=0.47\textwidth]{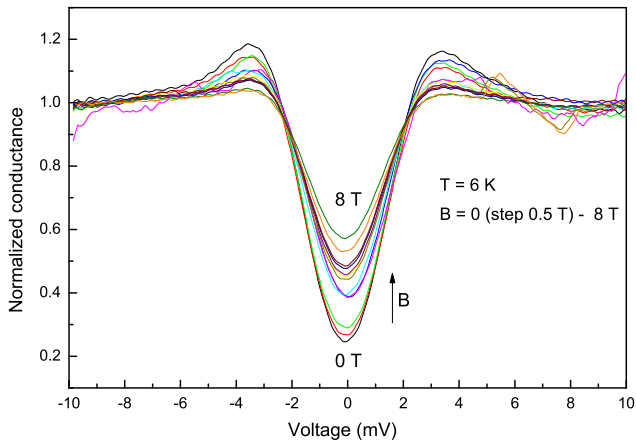}
\caption{Differential tunneling conductance of the NbN-Au tunnel junction measured at 6~K
	in magnetic fields up to 8~T with a 0.5~T step.} \label{ZFB}
\end{figure}

Tunneling measurements in magnetic fields up to 8\,T were performed at different temperatures.
A typical example of the measured data is shown in Fig.~\ref{ZFB} with the 
spectra taken at 6\,K in fields from 0 to 8 T with a 0.5\,T step. As it can be seen from the position of the peaks, the  superconducting gap is only slightly altered by the magnetic field and the
normalized conductance at zero bias increases with increasing magnetic field.
The Au STM tip was pressed against the sample surface to overcome the surface 
oxide layers. We deduced from imprints observed in an optical
microscope after the STM experiments  that the STM junction was of sub-micron size, in contrast to the
nanometer sizes which are typical for usual STM experiments. In such a situation when the magnetic field $B>B_{c1}$ is applied perpendicularly to the junction's interface on the NbN film, not only the film but also the junction area is in the mixed state comprising the Abrikosov vortices. Then, the measured tunneling conductance corresponds to an in-plane average
(over the junction's area) of the local densities of states. The zero-bias tunneling conductance is a measure of the averaged quasiparticle density of states at the Fermi level. In the first approximation the vortex cores represent the normal state areas while the rest of the junction is superconducting with a fully developed  superconducting order parameter. The increasing zero-bias tunneling conductance is thus proportional to the increasing volume fraction of vortex cores which shows a linear dependence on the applied magnetic field ($f_n=V_n/V\propto B/B_{c2}(T)$). In the paper of Samuely et al~\cite{Samuely1998}  it was shown that the measurement of normalized zero-bias tunneling conductance as a function of the field
strength is a very sensitive method to determine the upper critical magnetic field
$B_{c2}(T)$ for which the sample enters the normal state. A substantial advantage of this method is that even
if the value of $B_{c2}(T)$ exceeds the maximum attainable field, the 
upper critical field at a given temperature can be reliably determined  by 
a linear
 extrapolation of the zero-bias tunneling conductance to its normal state value.

\begin{figure}
 \center
\includegraphics [width=0.47\textwidth]{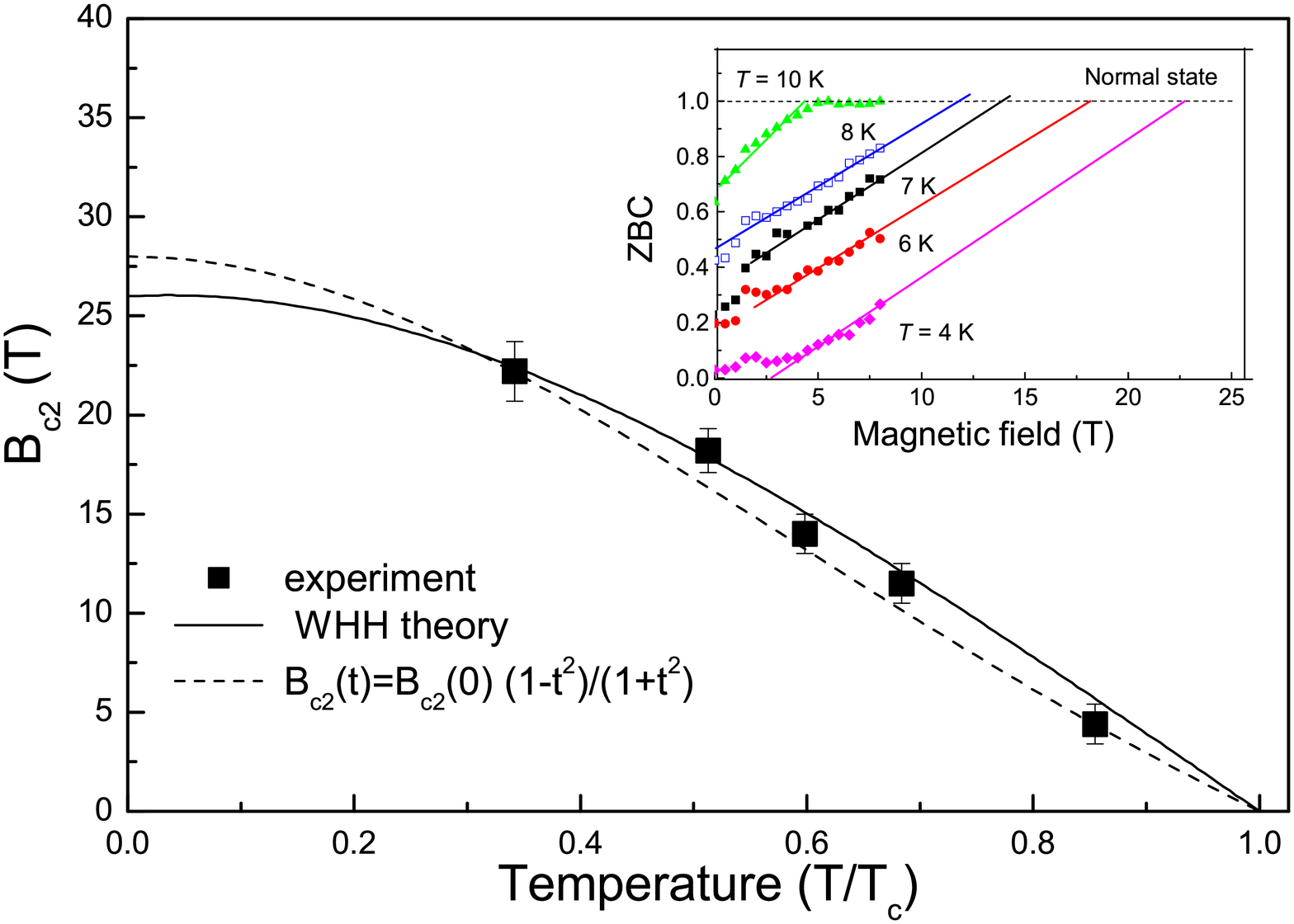}
\caption{Temperature dependence of the upper critical field $B_{c2}$. Symbols
 are data obtained from a linear extrapolation of the measured normalized
zero-bias tunneling conductance (ZBC) as a function of magnetic field shown in the inset.
The solid and dashed lines correspond to WHH and Tinkham models.
} \label{Bc2}
\end{figure}

The linear extrapolation of the normalized zero-bias tunneling conductance 
versus $B$ is shown in the inset of Fig.~\ref{Bc2}. In this way, $B_{c2}(T)$ was
determined for several temperatures and the results of this analysis are shown in the
main part of the figure. Although they were obtained only in a limited 
temperature range, the $B_{c2}$ values reveal a linear increase upon 
decreasing temperature below $T_c$ with a tendency to saturation below $0.5 
T/T_c$. Due to a lower number of data points two different theoretical model $B_{c2}(T)$ curves 
were drawn in order to extrapolate the zero-temperature value of the upper 
critical field. The classical Werthammer-Helfand-Hohenberg (WHH) dependence 
\cite{WHHmodel} is shown by the solid line yielding a correct fit. The extrapolated value of $B_{c2}(0)=26$\,T is found. Another frequently used model  developed by Tinkham
\cite{tinkham}, i.e. $B_{c2}=B_{c2}(0) (1-t^2)/(1+t^2)$, where $t=T/T_{c}(B=0)$, 
provides a fit of 
almost the same quality (dashed line) with a slightly higher value 
$B_{c2}(0)=28$\,T. With these $B_{c2}(0)$ estimates one derives the zero-temperature coherence length  $\xi(0)=3.5\pm 0.1$~nm.

\begin{figure}
 \center
\includegraphics [width=0.5\textwidth]{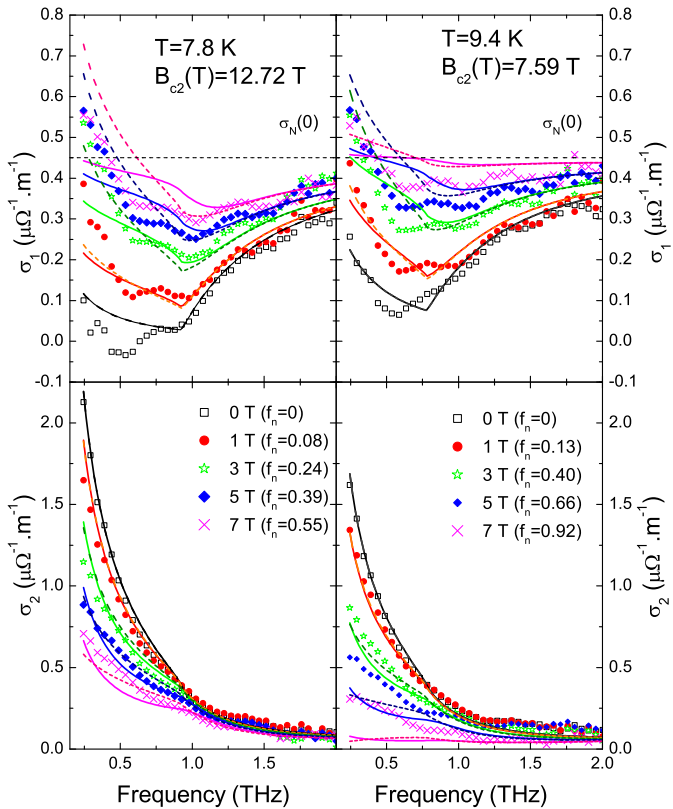}
\caption{Experimental data are compared with two different effective medium models:
Maxwell-Garnett theory (solid lines) and Bruggeman or effective medium approximation (dashed
lines). Following parameters were used in calculations: $\sigma_n(0)=0.45\, \mu\Omega^{-1}$.m$^{-1}$, $\tau=8$~fs, $T_c=12$~K and $2\Delta(0)/k_B T_c=4.25$. }
\label{MGTEMAgraph}
\end{figure}

We measured THz complex conductivity spectra in the Faraday geometry ($B$ perpendicular
to the film) for magnetic fields up to 7~T and at several temperatures starting
from 7.8~K up to $T_c$. The typical results are shown in Fig.~\ref{MGTEMAgraph}; at other
temperatures the spectra are qualitatively the same. The values of $\sigma_2$ in the THz
range slowly decrease with magnetic field while the frequency dependence of $\sigma_1$ is
more complex. At low THz frequencies $\sigma_1(\omega)$ grows with increasing 
magnetic field and exceeds the normal state value $\sigma_n(0)$. As the upper critical field is approached, $\sigma_1$
starts to decrease  and it reaches to the normal-state value 
$\sigma_n(0)$ at $B_{c2}(T)$. This is difficult to see at $T=9.4$~K (Fig.~\ref{MGTEMAgraph}) but it was clearly observed at $T=10$~K (not shown). For frequencies above
the optical gap $\sigma_1$ monotonically increases towards its normal-state value.

\section{Theoretical models}
Various aspects of the high-frequency response of Abrikosov state are described by effective medium models, e.g. the Maxwell-Garnet or the Bruggeman theories, and by the Coffey-Clem
model. These two approaches are fundamentally different. The effective medium 
theories treat vortex cores as normal-state inclusions embedded in a superconducting matrix.
 The induced depolarization fields  
modify the macroscopic response of the sample. 
These models do not take into account effects due to the vortex motion or linked 
to the regular arrangement of
 vortices into a hexagonal lattice (a random distribution is assumed instead).
Effects caused by vortex dynamics are described by the Coffey-Clem model which  assumes that the whole volume is filled
with a mixture of a normal fluid and a superfluid whose fractions depend on the
temperature and the external magnetic field; vortices are introduced only as zero-volume lines. 
Let us note that the applicability of the Coffey-Clem model
is limited to frequencies below the superconducting gap.

For low temperatures Ikebe et al. \cite{Ikebe2009} made an attempt to combine the Coffey-Clem model with the Maxwell-Garnet theory. 
The generalization of this approach for finite temperatures with a substantial number of quasiparticles present, however,
is not clear.

\subsection{Effective medium models}
As mentioned above, the superconducting film is considered as a system consisting of
 a superconducting matrix and cylindrical inclusions of the 
 normal-state material representing vortex cores \cite{Clemmodel}.
 The radius of the vortex core is defined by the coherence length $\xi$,
i.e., it is of the order of a few nanometers; the inter-vortex distance is usually of the
same order for fields applied in our experiment. The typical wavelengths of the far-infrared radiation (hundreds of micrometers)
are much larger, therefore the electromagnetic radiation cannot sense individual
vortices and an effective complex conductivity can be used to describe the
macroscopic properties of the system. The upper frequency limit of the effective medium models is given merely by the size of the 
vortices.

The Maxwell-Garnet theory~\cite{MGT_original_paper} (MGT) is suitable for dilute systems of
inclusions (vortices) with a percolated matrix (superconducting
state) and in the case of cylindrical inclusions it takes the form~\cite{Carr}:
\begin{equation}
\label{mgt}
 \tilde{\sigma}_{\mathrm{MG}}=\frac{2 f_n \tilde{\sigma}_s (\tilde{\sigma}_n - \tilde{\sigma}_s)}{(1-f_n)(\tilde{\sigma}_n - \tilde{\sigma}_s)+2\tilde{\sigma}_s}+\tilde{\sigma}_s,
\end{equation}
where $f_n=V_n/V$ is the volume fraction of vortex cores, $\tilde{\sigma}_s$ and
$\tilde{\sigma}_n$ are complex conductivities of the superconducting and normal state,
respectively. Recently, Rychetsky \cite{Rychetsky2004} argued that the MGT formula holds even
 for high concentrations of inclusions as long as the matrix is percolated.

The Bruggeman approach~\cite{Bruggeman} assumes that both superconducting and 
vortex-core components are surrounded by an effective medium with the effective permittivity
$\tilde{\varepsilon}_{\mathrm{B}}$. Local field and effective conductivity values are found in
the self-consistent way and one finds~\cite{Carr}
\begin{equation}
\label{ema} \tilde{\sigma}_{\mathrm{B}}=\frac{1}{2} \left (\beta \pm
\sqrt{\beta^2+4\tilde{\sigma}_n \tilde{\sigma}_{s}} \right ).
\end{equation}
Here $\beta=(1 - 2f_n) \left( \tilde{\sigma}_n \ - \tilde{\sigma}_{s} \right)$ and the
sign is chosen so that the physically relevant solution $\sigma_{1,\mathrm{B}}\geq0$ is
obtained.

\subsection{Coffey-Clem model}

Coffey and Clem \cite{Coffey-Clem_model_1991} solved self-consistently the modified
London equation in the presence of vortices and found that the vortex motion can modify
both the dissipative and the inductive part of the complex conductivity. Explicitly,
their model can be expressed as:
\begin{equation}
 \tilde{\sigma}_{\mathrm{CC}}=\frac{(1-f)\tilde{\sigma}_s+f \tilde{\sigma}_n}{1+(1-f)  \tilde{\sigma}_s/\tilde{\sigma}_{vd}},
\label{CCeq}
\end{equation}
where $f=1-(1-T^4/T_c^4)(1-B/B_{c2}(T))$ is the normal-fluid fraction and
$\tilde{\sigma}_{vd}$ is the contribution to the conductivity induced by the vortex
dynamics \cite{PompeoSilva2008}:
\begin{equation}
 \tilde{\sigma}_{vd}=\frac{\sigma_n}{b}\left( \frac{1+i\omega/\omega_0}{\epsilon+i\omega/\omega_0}\right),
\label{vortex_resistivity}
\end{equation}
where the dimensionless parameter $\epsilon$ varies between 0 (at zero temperature) and 1
(at $T_c$) and describes effects of the flux creep, $\omega_0$ is an effective
depinning frequency and $b=B/B_{c2}(T)$. Depending on temperature, on the magnetic field,
and on the ratio $\omega/\omega_0$, three different regimes emerge
\cite{Golosovsky_vortex_dynamics_overview}: a flux-creep regime (where the flux creep
dominates and the response is mainly dissipative), a pinning regime (where the inductive
response prevails) and a flux-flow regime (where vortices move collectively and the
dissipative response dominates).

\section{Discussion}
\subsection{Sources of errors}

An important source of errors in thin film measurements come from the uncertainty in the substrate thickness which significantly influences the phase of the transmitted THz wave~\cite{kadlecch}. The error bars shown in Fig. 2 take into account two effects: the statistical error in the time-domain THz wave form calculated from individual data accumulations and a systematic error due to a substrate thickness uncertainty of 1~$\mu$m and can be considered as typical for all other data shown in this paper. 

The oxidized layer covering the NbN film influences the evaluation 
of the complex conductivity. On the one hand, the presence of an ultrathin niobium
oxide layer does not have any influence on the THz transmission of the sample. On the other hand, if such a layer develops,
the superconducting film thickness is necessarily reduced by  $\Delta d_{film}$. This is a source of error
in the evaluated conductivity given as  $\Delta \sigma_{sc}= (\Delta d_{film}/d_{film}) \, \tilde{\sigma}_{sc}$. In our case it may be responsible for an error of up to 9\%.  However, this error  only  rescales
$\sigma_n(0)$ so that the qualitative analysis is not affected. Nevertheless this error in principle should be added to the error bars shown in Fig.~\ref{Zimmerman}.

The in-plane inhomogeneity leading to the observed 
differences in the values of $T_c$ can cause some minor variations of the 
complex THz conductivity.  As far as the sample temperature is 
sufficiently far from $T_c$ the inhomogeneity will therefore not play a significant role.

Finally, small systematic oscillations observed in the spectra of Fig.~\ref{Zimmerman} are probably due to a very small error in the instrumental function (parasitic reflection in the waveform) which can lead to a visible Fabry-Pérot-like effect in the conductivity of a thin film.

In zero magnetic field the value of the superconducting gap was determined from the
normalized tunneling conductance. The ratio $2\Delta(0)/k_B T_c$ was observed to vary
from 4.25 to 4.5 along the surface of the sample; these values are slightly higher than
what is typically observed~\cite{Uwe,Ikebe2009,Matsunaga2012,Tesar2011,Pambianchi1994}, nevertheless even higher
values were also found~\cite{Beck2011}. These values suggest NbN is a strong-coupling superconductor and that, in general,
high-frequency (THz) response should be treated within the framework of Eliashberg formulas following Nam~\cite{Nam1967I,Nam1967II}. The dominant effect of strong coupling is a uniform decrease of $\sigma_2(\omega)$ while the effect on $\sigma_1(\omega)$ is only minor. Nevertheless, based on previous good experience with analysis of similar data without accounting for the strong-coupling effects \cite{Uwe,Karecki,Kang2011,Matsunaga2012,Beck2011} we believe that these effects are not substantial and that the BCS theory based Zimmermann model \cite{Zimmmermann}  describes THz properties of NbN films adequately.

\subsection{Modelling the experimental data}
\begin{figure}
 \center
\includegraphics [width=0.5\textwidth]{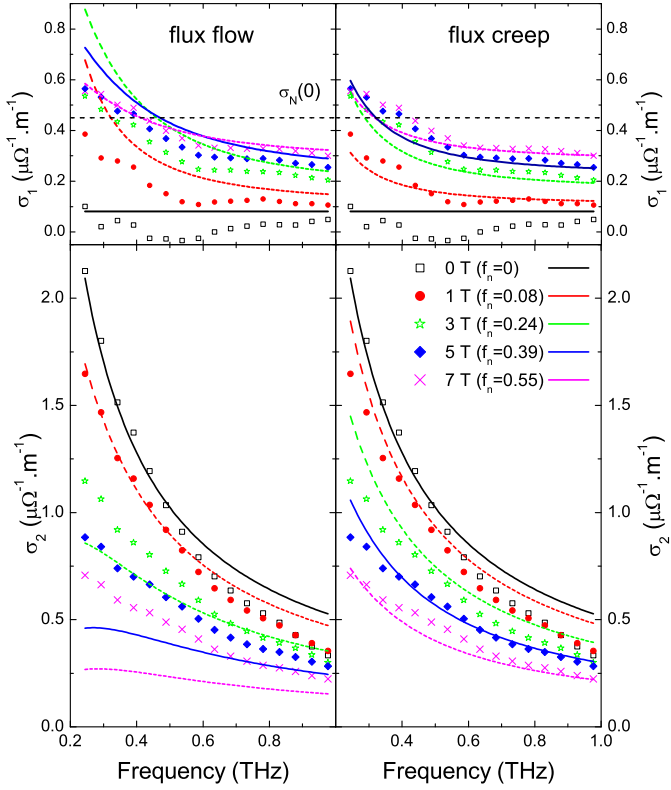}
\caption{ Coffey-Clem model in two different limits: flux flow (left panel) and thermal
creep (right panel) for $T=7.8~K$. Experimental data are plotted as a points. }
\label{CCgraph1}
\end{figure}

\begin{figure}
 \center
\includegraphics [width=0.5\textwidth]{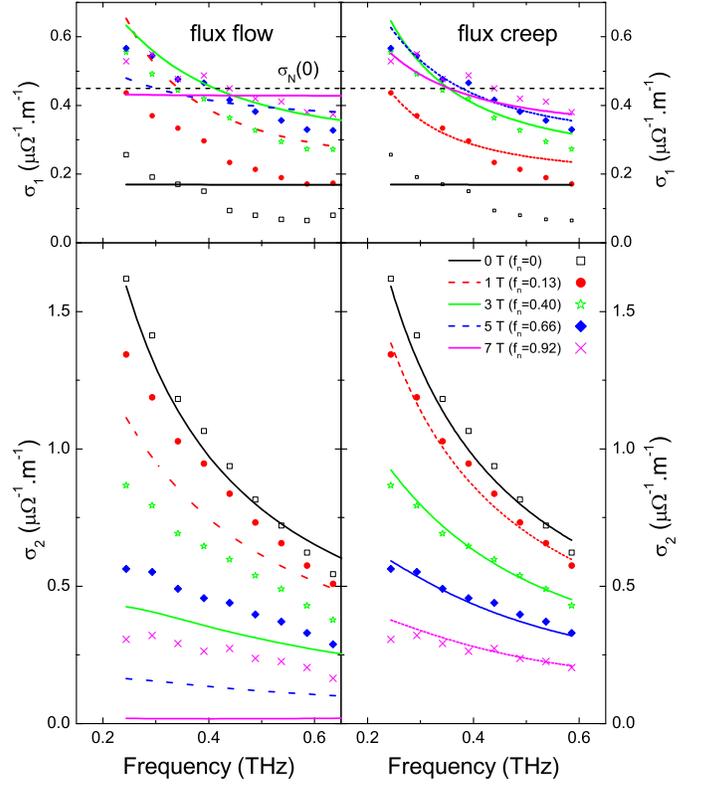}
\caption{ Coffey-Clem model in two different limits: flux flow (left panel) and thermal
creep (right panel) for $T=9.4~K$. Experimental data are plotted as a points. }
\label{CCgraph2}
\end{figure}

The complex conductivity of the normal state of our NbN sample is well described by the Drude
model: it is almost purely real and frequency-independent in the observed range, the relaxation time $\tau$ thus  reaches a
value of a few femtoseconds only and its precise value cannot be established 
reliably. Below $T_c$  and 
in the absence of magnetic field, the experimental data are well 
described by the
Zimmermann model~\cite{Zimmmermann}, see Fig.~\ref{Zimmerman}. The two-fluid model~\cite{tinkham} fails
to describe the features connected with the gap which are clearly observed in the real part of
the conductivity spectra; by contrast, it describes the imaginary part fairly well,
see the zero-field data  shown in Figs.~\ref{CCgraph1} and \ref{CCgraph2}.

In the analysis of the transmission measurements in magnetic field we compare our experimental
results with the theoretical models presented in the previous section. Almost all the parameters of
these models were determined using STS or from the zero-field
transmission measurements. In the frame of Coffey-Clem model only the creep parameter $\epsilon$ and the effective depinning frequency
$\omega_0$ are free parameters.

Let us start with an application of the effective medium theories, since the 
local field effects (almost negligible in a microwave range where 
$|\sigma_s|\gg|\sigma_n|$) become important at terahertz frequencies. In Figure 
\ref{MGTEMAgraph}, the predictions by the Maxwell-Garnett and the Bruggeman theories are compared
with experimental data for two different
temperatures. For $T=9.4$~K and B varying from 0 to 7 T, the volume fraction of vortex cores $f_n=B/B_{c2}(T)$ spans
almost over the entire interval from 0 to 1 (normal state) providing thus 
an excellent opportunity to test these theories.
On the one hand, both theories give similar results and agree with experimental data for the frequencies above the gap and also for low
magnetic fields. On the other hand,
at low frequencies and high magnetic fields, both theories fail to describe 
the experimental values of $\sigma_2(\omega)$. Furthermore, in the frame 
of the Maxwell-Garnett theory, the real part of $\sigma_1$ is limited by $\sigma_n(0)$ which is in disagreement with the experimental data. In our analysis, we neglected that conductivity of the  superconducting matrix $\tilde{\sigma}_s$ is modified
by the magnetic field \cite{Skalski1964,Xi2010}. 
Xi et al.~\cite{Xi2013} took this effect into account, but their 
numerical calculations lead to a further decrease in $\sigma_2$, which disagrees with our experimental data. We thus concluded that this effect cannot explain the discrepancies between  the predictions of the effective medium models and the experimental data. 

We agree with the argument of Xi et al.~\cite{Xi2013} that the 
topology assumed by the Bruggeman theory does not match that of the 
vortex state. From this point of view one may expect that the MGT 
 should be more suitable. In our previous 
study~\cite{Sindler2010}, however, we found that our experimental data 
were both quantitatively and qualitatively better described by the 
Bruggeman theory. This theory accidentally mimics the effect of vortex dynamics at low 
frequencies and high magnetic fields as we can assume on the basis of the present study.

In Figures~\ref{CCgraph1} and \ref{CCgraph2} the low frequency data ($\hbar \omega<2\Delta$) 
are compared with the Coffey-Clem model, Eq.~\ref{CCeq}. Since this model is not adequate for higher frequencies,
conductivity of higher frequencies is not shown. Due to the weak dependence of the
complex conductivity on free parameters, $\epsilon$ and $\omega_0$ can not be reliably
determined from our data. Therefore we compare our data with two limiting 
cases--- the flux-flow regime (where $\omega \gg \omega_0$ and the flux creep 
can be neglected, see panel A) and the flux-creep regime (where $\omega \ll 
\omega_0$ and the flux creep dominates, see panel B). Our data (both real and 
imaginary parts of $\tilde{\sigma}_{sc}$) suggest that the 
flux flow does not occur, which is in disagreement with previous observations \cite{Ikebe2009,Rosenblum1964}. However, the Coffey-Clem model in the case of the flux creep provides an excellent description of $\sigma_2$ and reasonable description of $\sigma_1$.

\section{Summary}

We report properties of a thin NbN film deposited on a 
high-resistivity Si substrate, obtained by time-domain
THz  and scanning tunneling spectroscopies. STS revealed 
the presence of a continuous oxide layer
on the top of the film. We found that the optical gap $2\Delta(0)$ is varying between $2.1-2.3$ meV and 
the critical temperature $T_c$ between 11.5 and 12~K. This indicates a strong coupling superconductivity with 
$2\Delta(0)/k_B T_c=4.25-4.5$.

The values of the upper critical magnetic field $B_{c2}$ were determined 
by a linear extrapolation
 of the normalized zero-bias tunneling conductance as a function of the field strength. Its temperature
 dependence is well described by the WHH model with $B_{c2}(0)=26$ 
 T. The measured zero-field THz complex 
conductivity is well described by Zimmermann model \cite{Zimmmermann}. 

For low magnetic fields we found that both the Maxwell-Garnett and Bruggeman theories  give similar results and agree
 with the experimental data. At high magnetic fields, however, they both fail to  describe the dissipation at low frequencies
and the inductive response, $\sigma_2$. The obvious reason is that they do not account for the effects of vortex dynamics, which are  particularly important at low THz frequencies. Fortunately, the Coffey-Clem two-fluid approach which neglects pair-breaking process for frequencies above the optical gap, can be used for low THz frequencies. Our experimental results were compared  with  the Coffey-Clem model assuming either the flux-flow or the flux-creep regime. Based on the excellent description of $\sigma_2$ and reasonable description of $\sigma_1$ we concluded that the Coffey-Clem model in the flux-creep regime describes best the THz properties of our NbN sample.

\section{Acknowledgment}
We are grateful to K.~Ilin and M.~Siegel for preparing and characterizing the NbN sample.
This work was supported by the following projects: GA\v{C}R under contract P204/11/0015, COST action MP1201, CFNT MVEP-the Centre of Excellence of the Slovak Academy of Sciences, FP7 
MNT—ERA.Net II, ESO, the EU ERDF (European Union Regional Development Fund) grant No. ITMS26220120005, VEGA No. 2/0135/13 and
 the APVV-0036-11 grant of the Slovak R\&D Agency. The liquid nitrogen for the experiment was sponsored by US Steel Ko\v{s}ice, s.r.o.

\end{document}